\documentclass[twocolumn,preprintnumbers,amsmath,amssymb,superscriptaddress]{revtex4}
\usepackage{graphicx}
\usepackage{dcolumn}
\usepackage{bm}
\usepackage{soul}
\usepackage{color}
\usepackage{epstopdf}
\usepackage[version=3]{mhchem}
\usepackage{lipsum}
\usepackage[outercaption]{sidecap}
\usepackage{floatrow}

\begin{document}

\preprint{APS/123-QED}

\title{Dynamic Gradients, Mobile Layers, \ce{T_g} Shifts, Role of Vitrification Criterion and Inhomogeneous Decoupling in Free-Standing Polymer Films }
\author{Anh D. Phan}
\affiliation{Department of Physics, University of Illinois, 1110 West Green St, Urbana, Illinois 61801, USA}
\author{Kenneth S. Schweizer}
\affiliation{Department of Materials Science and Chemistry, Frederick Seitz Materials Research Lab, University of Illinois at Urbana-Champaign}
\email{kschweiz@illinois.edu}

\date{\today}

\begin{abstract}
The force-level Elastically Collective Nonlinear Langevin Equation (ECNLE) theory of activated relaxation in glass-forming free standing thin films is re-visited to improve its treatment of collective elasticity effects. The naive cut off of the isotropic bulk displacement field approximation is improved to explicitly include some aspects of spatial anisotropy with a modified boundary condition consistent with a step function liquid-vapor interface. The consequences of this improvement on dynamical predictions are quantitative but of significant magnitude, and in the direction of further speeding up dynamics and further suppressing \ce{T_g}. The theory is applied to thin films and also (for the first time) semi-infinite thick films to address qualitatively new questions for three different polymers of different dynamic fragility. Variation of the vitrification time scale criterion over many orders of magnitude is found to have a surprisingly minor effect on changes of the film-averaged \ce{T_g} relative to its bulk value. The mobile layer length scale grows strongly with cooling, and correlates in a nearly linear manner with the effective barrier deduced from the corresponding bulk isotropic liquid alpha relaxation time. The theory predicts a new type of spatially inhomogeneous "dynamic decoupling" corresponding to an effective factorization of the total barrier into its bulk temperature-dependent value multiplied by a function that only depends on location in the film. The effective decoupling exponent grows monotonically as the vapor surface is approached. Larger reductions of the absolute value of Tg shifts in thin polymer films are predicted for longer time vitrification criteria and more fragile polymers. Quantitative no-fit-parameter comparisons with experiment and simulation for film-thickness-dependent \ce{T_g} shifts of polystyrene and polycarbonate (PC) are in reasonable accord with the theory, including a nearly 100 Kelvin suppression of \ce{T_g} in 4 nm PC films. Predictions are made for polyisobutylene thin films. 
\end{abstract}

\pacs{Valid PACS appear here}
\maketitle

\section{Introduction}
Activated glassy dynamics, mechanical properties and vitrification in thin polymer films with diverse boundary conditions are problems of great scientific interest \cite{1,2,3,4,5} and also are of importance in applications for sensors \cite{6}, photoresists \cite{7}, coatings \cite{8} and optoelectronic devices \cite{9}. Despite much effort over the past two decades, including the construction of many different phenomenological models built on different ansatzes \cite{2,5,10,11,12,13,14,15,16,17,18}, the physical mechanisms responsible for the observed phenomena remain not very well understood. Broadly speaking, this reflects the complexity of activated relaxation in bulk liquids in addition to the major complications of confinement, interfaces and spatial inhomogeneity. 

Free standing thin films with two vapor interfaces, or semi-infinite thick films with one vapor interface, are perhaps the simplest realization of confined films. They would seem to be the most likely candidates for realizing some (perhaps limited) universality of behavior. Extensive experimental \cite{1,2,3,19,20,21,22,23,24,25,26,27,28} and simulation \cite{2,5,14,29,30,31,32} efforts have revealed that the structural relaxation time in these systems can speed up enormously and mobile layers extend rather deeply into the film, with corresponding large film-averaged reductions of the glass transition temperature, \ce{T_g}. 

Recently, Mirigian and Schweizer proposed a theory for free standing films built on a quantitative, force-level statistical mechanical theory of alpha relaxation in isotropic bulk supercooled molecular \cite{33,34,35} and polymer liquids \cite{36,37} the “Elastically Collective Nonlinear Langevin Equation” (ECNLE) theory. Quantitative tractability for real materials is achieved based on an a priori mapping of chemical complexity \cite{34} to a thermodynamic-state-dependent effective hard sphere fluid. The structural relaxation event involves coupled cage-scale hopping and a longer range collective elastic distortion of the surrounding liquid, resulting in two inter-related, but distinct, barriers. The theory accurately captures the key features of the alpha time of molecular liquids over 14 decades \cite{33,34,35}. Generalization to polymer liquids is based on a primitive disconnected Kuhn segment model \cite{36,37}.

The extension of ECNLE theory to free-standing films predicts the spatial gradient of the alpha relaxation time as a function of temperature, film thickness and location in the film, and from this Tg gradients, film-averaged Tg shifts, and other properties can be deduced \cite{38,39,40}. Relaxation speeds up for purely dynamical reasons via reduction of the cage barrier near the liquid-vapor interface due to loss of neighbors, and reduction of the collective elasticity cost for hopping due to a cut off mechanism which is operative even for re-arrangements deep in the film. The theoretical results have been encouragingly compared with experiment. However, the daunting complexity of the spatially inhomogeneous activated dynamics problem in polymer thin films encourages attempts to improve the theory. Here, we re-visit the prior naive treatment of collective elasticity based on a cutoff of the isotropic bulk displacement field model \cite{38,39,40} to include some aspects of the spatial anisotropy of the elastic field and a modified boundary condition at the vapor interface. 

After briefly reviewing the prior version of bulk and thin film ECNLE theory in section II, the improved treatment is derived in section III. Representative calculations of the elastic displacement field and its consequences on barrier gradients are presented. This improved formulation is then used to examine dynamical effects in thin films and also (for the first time within the ECNLE approach) one-interface semi-infinite thick films. Qualitatively new questions not previously addressed are studied. Section IV focuses on the simpler case of thick films, and examines three fundamental issues. (i) Spatial gradients of the alpha time and local \ce{T_g} of polystyrene over a wide range of temperature and for different vitrification criteria. (ii) Mobile layer length scale, its evolution with temperature, and whether it is related to the bulk alpha relaxation time. (iii) Does a unique form of spatially inhomogeneous "dynamic decoupling" occurs in films? Section V considers free standing thin films and analyzes how the total barrier as a function of spatial location and film thickness is modified by our new treatment of the elastic displacement field. These results are applied to compute pseudo-thermodynamic and dynamic measures of the film-averaged \ce{T_g} shifts, the influence of vitrification timescale criterion, and polymer chemistry effects with an emphasis on the influence of variable bulk fragility. Several theoretical results are quantitatively compared to simulations and experiments on polystyrene and polycarbonate, and predictions are made for polyisobutylene. The article concludes in Section VI with a discussion.

\section{ECNLE Theory of Bulk Liquids and Thin Films}
As relevant background we briefly review the present state of ECNLE theory for bulk liquids and free standing thin films, along with the mapping of chemical complexity to effective hard sphere fluids. All aspects have been discussed in great detail in prior papers \cite{33,34,35,36,37,38,39,40}.
\subsection{Bulk Liquids}
\begin{figure}[htp]
\includegraphics[width=8cm]{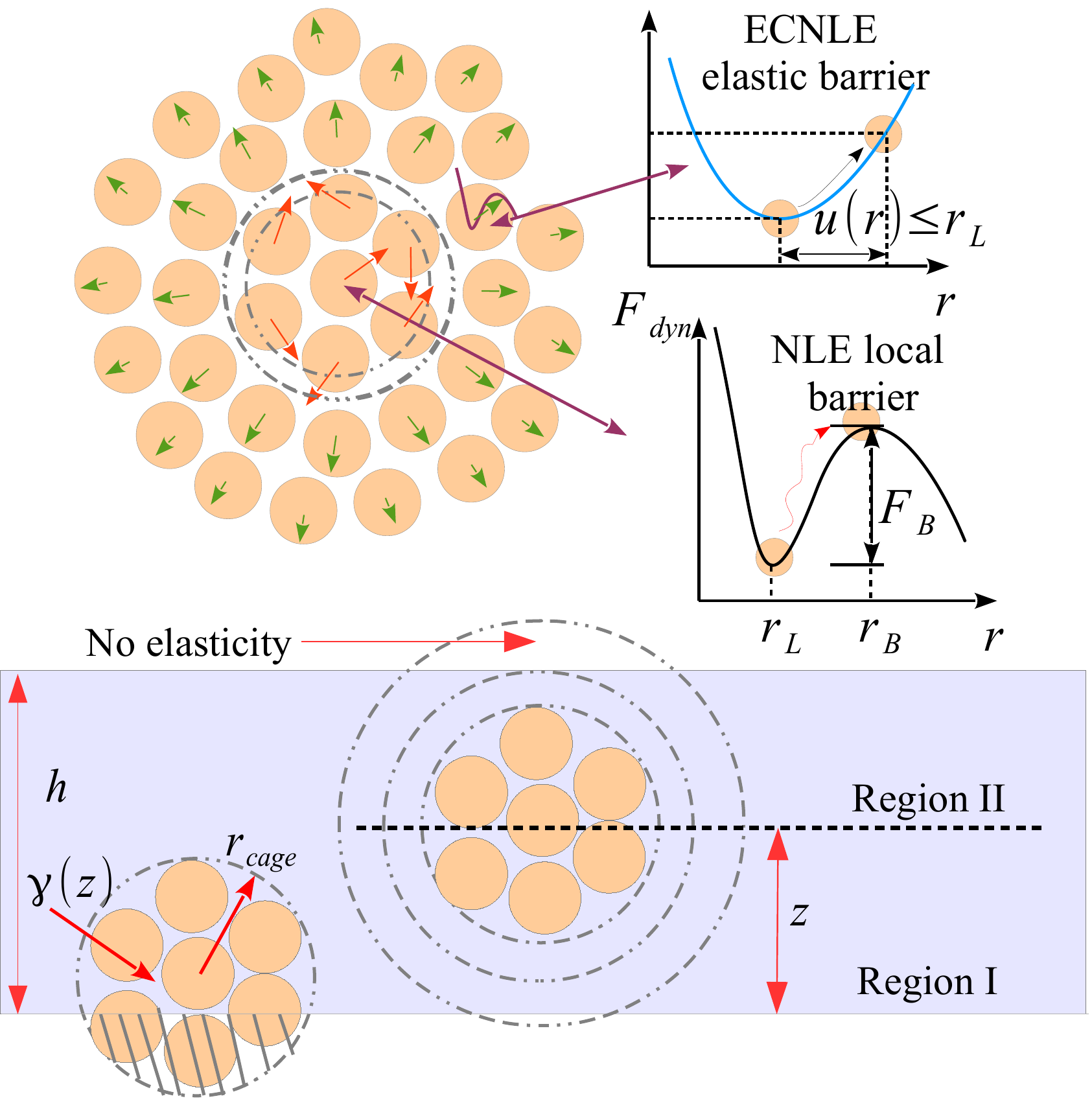}
\caption{\label{fig:1}(Color online) Upper panel: schematic illustration of the conceptual elements of ECNLE theory for bulk spherical particle fluids. The dynamic free energy and important length and energy scales are indicated. Lower panel: schematic illustration of free standing films indicating the coordinate z, film thickness h, regions I and II, cage radius, the sharp interface, loss of neighbors near the surface, and cut off of the elastic displacement field at the vapor interface.}
\end{figure}

Consider a one-component liquid of spherical particles (diameter, $d$) of packing fraction $\Phi$. ECNLE theory describes the activated relaxation of a tagged particle as a mixed local-nonlocal rare hopping event \cite{33}. Figure \ref{fig:1} shows a cartoon of the key physical elements. The foundational quantity is an angularly-averaged displacement-dependent dynamic free energy, $F_{dyn}(r)=F_{ideal}(r) + F_{caging}(r)$, the derivative of which is the effective force on a moving tagged particle in a stochastic NLE:
\begin{eqnarray}
\frac{F_{dyn}(r)}{k_BT}&=&-3\ln\left(\frac{r}{d}\right)-\rho\int\frac{d\mathbf{q}}{(2\pi)^3}\frac{S(q)C^2(q)}{1+S^{-1}(q)}\nonumber\\
&\times&\exp\left[-\frac{q^2r^2}{6}\left( 1+S^{-1}(q)\right) \right],
\label{eq:1}
\end{eqnarray}
where $k_B$ is Boltzmann’s constant, $T$ is temperature, $\rho$ is the number density, $r$ is the displacement of the particle from its initial position, $S(q)$ is the static structure factor, $q$ is wavevector, and $C(q)=\left[ 1-S^{-1}(q)\right]/\rho$ is the direct correlation function \cite{41}. The leading term of Eq.(\ref{eq:1}) favors the fluid state, and the second term corresponds to a trapping potential due to interparticle forces which favors localization. As the density (or temperature) of the liquid exceeds (goes below) a critical value, a local barrier $F_B$ in $F_{dyn}(r)$ emerges signaling transient localization by neighboring cage particles where the cage radius $r_{cage}\approx 1.3-1.5d$ is defined as the position of the first minimum in the radial distribution function $g(r)$. Figure \ref{fig:1} also defines a transient localization length $r_L$, barrier position $r_B$, and jump distance $\Delta r = r_B-r_L$; the latter is $\approx 0.2-0.35d$ for hard sphere fluids.

Large amplitude local hopping is strongly coupled to a spatially long-range collective elastic adjustment of all particles outside the cage required to create the small amount of extra space to accommodate a hop. Determination of the precise form of the elastic displacement field is a difficult inverse problem. As a technical approximation, the liquid outside the cage is treated as a continuum elastic material which allows calculation of the displacement field, $u(r)$, using standard linear continuum mechanics \cite{42} supplemented by a microscopic boundary condition. The field obeys:
\begin{eqnarray}
\left(K + \frac{G}{3} \right)\nabla(\nabla \mathbf{u}) + G\nabla^2\mathbf{u} = 0,
\label{eq:2}
\end{eqnarray}
where $K$ and $G$ are the bulk and dynamic shear modulus, respectively. The radially-symmetric solution is a scale-free displacement field which decays as an inverse square power law \cite{33}:
\begin{eqnarray}
u(r)=\Delta r_{eff}\frac{r_{cage}^2}{r^2}, \qquad r\geq r_{cage}
\label{eq:3}
\end{eqnarray}
The amplitude is set by the small orientationally-averaged mean cage expansion length, $\Delta r_{eff}$ \cite{33}:
\begin{eqnarray}
\Delta r_{eff}\approx 3\Delta r^2/32r_{cage} \leq r_L.
\label{eq:4}
\end{eqnarray}
The prefactor $3/3$2 follows from assuming each spherical particle in the cage independently hops in a random direction by $\Delta r$, which is in the spirit of the dynamic mean field nature of NLE theory \cite{43}.

Since in ECNLE theory the local and long range elastic aspects are intimately related, the elastic barrier is determined in a "molecular Einstein-like" manner by summing over all harmonic particle displacements outside the cage region thereby yielding \cite{33}:
\begin{eqnarray}
F_{elastic} &=& \rho\frac{K_0}{2}\int_{r_{cage}}^\infty dr 4\pi r^2 u^2(r)g(r)\nonumber\\
&\approx& 12K_0\Phi\Delta r_{eff}^2\left(\frac{r_{cage}}{d}\right)^3,
\label{eq:5}
\end{eqnarray}
where $r$ is relative to the cage center and $K_0=3k_BT/r_L^2$ is the curvature (harmonic stiffness) of the dynamic free energy at its minimum. We note the long range nature of the integrand in Eq.(\ref{eq:5}) which decays as $\sim r^{-2}$, and hence the total elastic barrier converges slowly to its full value as $\sim r^{-1}$. This aspect will also be important in thin films or near an interface.

The sum of the coupled (and in general temperature and density dependent) local and elastic collective barriers determine the total barrier for the alpha process:
\begin{eqnarray}
F_{total} = F_B + F_{elastic}.
\label{eq:6}
\end{eqnarray}
The elastic barrier increases much more strongly with increasing density or cooling than its cage analog, and dominates alpha relaxation time growth as the laboratory glass transition is approached.33,34 A generic measure of the structural relaxation time follows from a Kramers calculation of the mean first passage time over the barrier. For barriers in excess of a few thermal energy units one has:
\begin{eqnarray}
\frac{\tau_\alpha}{\tau_s} = 1+ \frac{2\pi(k_BT/d^2)}{\sqrt{K_0K_B}}\exp{\frac{F_B+F_{elastic}}{k_BT}},
\label{eq:7}
\end{eqnarray}
where $K_B$ is the absolute magnitude of the barrier curvature. The alpha time is expressed in units of a "short time/length scale" relaxation process (cage-renormalized Enskog theory), $\tau_s$, the explicit formula for which is given elsewhere \cite{34}. Physically, it captures the alpha process in the absence of strong caging defined by the parameter regime where no barrier is predicted (e.g., $\Phi < 0.43$ for hard spheres \cite{43}). The latter condition corresponds to being below the naive mode coupling theory ideal glass transition \cite{43,44} which in ECNLE theory is manifested as a smooth dynamic crossover. 

The above theory can be applied to any fluid of spherical particles. It is rendered predictive for molecular liquids via a mapping \cite{34,36} to an effective hard sphere fluid guided by the requirement that it exactly reproduces the equilibrium dimensionless density fluctuation amplitude (compressibility) of the liquid, $S_0(T) = \rho k_BT\kappa_T$. This quantity sets the amplitude of nm-scale density fluctuations, and follows from the experimental equation-of-state (EOS). The mapping is \cite{34,36}: 
\begin{eqnarray}
S_0^{HS}&=&\frac{(1-\Phi)^4}{(1+2\Phi)^2}\equiv S_{0,exp}=\rho k_BT\kappa_T\nonumber\\
&\approx &\frac{1}{N_s}\left(-A+\frac{B}{T} \right)^{-2}.
\label{eq:8}
\end{eqnarray}
The first equality follows from adopting the Percus-Yevick (PY) integral equation theory \cite{41} for hard sphere fluids. The final approximate equality is an analytically derived form that accurately describes experimental data \cite{34}. This mapping determines a material-specific, temperature-dependent effective hard sphere packing fraction: $\Phi_{eff}(T;A,B,N_s)= 1 + \sqrt{S_0^{expt}(T)}-\sqrt{S_0^{expt}(T) + 3\sqrt{S_0^{expt}(T)}}$. In practice, four known chemically-specific parameters enter \cite{34}: $A$ and $B$ (interaction site level entropic and cohesive energy EOS parameters, respectively), the number of elementary sites that define a rigid molecule, $N_s$ (e.g., $N_s=6$ for benzene), and hard sphere diameter, $d$. Knowledge of $\Phi(T)$ allows $g(r)$ and $S(k)$ to be computed using PY theory, which determines $F_{dyn}(r)$, from which all dynamical results follow. With this mapping, ECNLE theory makes alpha time predictions for molecular liquids with no adjustable parameters which have been shown to be quantitatively accurate over 14 decades for nonpolar molecules and less accurate for hydrogen-bonding molecules \cite{34,35}.

Polymers have additional complexities associated with connectivity and conformational isomerism. As a minimalist model the liquid is replaced by a fluid of disconnected Kuhn segments modeled as non-interpenetrating hard spheres \cite{36}. Polymer-specific errors must be incurred, and a one-parameter non-universal version of ECNLE theory for polymer melts has been developed \cite{37} based on the hypothesis the amount of cage expansion depends on sub-nm chemical details coarse-grained over in the effective hard sphere description. Nonuniversality enters via a modified jump distance, $\Delta r \rightarrow \lambda\Delta r$, where the constant numerical factor $\lambda$ is adjusted to simultaneously provide the best theoretical description of both $T_g$ and fragility for a specific polymer chemistry. From Eqs. (\ref{eq:2}) and (\ref{eq:3}), this results in $F_{elastic}\rightarrow \lambda^4F_{elastic}$. The relative importance of the local versus collective elastic barrier thus acquires a polymer-specificity.

In this article, we present results for polystyrene (PS), polycarbonate (PC) and polyisobutlyene (PIB). All required parameters and mapping details are identical to those reported previously \cite{37}. Since PS is our "baseline" system, we recall the relevant parameters: $N_s(PS) = 38.4$, $d = 1.16$ nm, $A(PS)=0.618$, $B(PS)=1297$ and $\lambda_{PS} = 1$. The above three polymers were chosen since they have widely varying fragility, defined as $m=\left(\partial\log\tau_\alpha/\partial(T_g/T)\right)|_{T=T_g}$. Based on the bulk vitrification criterion of $\tau_\alpha(T_g)=100$ s, prior work \cite{37} found: PS has a fragility $m \sim 110$ with $\lambda_{PS} = 1$, PC has a very high fragility of $m \sim 140$ with $\lambda_{PC} = \sqrt{2}$, and PIB has a very low fragility of $m \sim 46$ with $\lambda_{PIB} = 0.47$.

\subsection{Free Standing Films}
Films with interfaces exhibit broken symmetry resulting in every material property (thermodynamic, structural, dynamic) becoming spatially heterogeneous and anisotropic. Treating all of this complexity theoretically is intractable, and hopefully not necessary. Thus, in the past \cite{38,39,40} a minimalist approach was adopted based on the hypothesis that the most important effects are purely dynamical. We assume no changes of thermodynamics or structure in the film. This ansatz is consistent with recent machine-learning based analysis of simulations of free standing films \cite{45} which found the large dynamical changes observed are not related to any change of equilibrium properties. For simplicity, we also adopt a step-function density profile in the direction orthogonal to the interface. 

There remains the challenge of formulating a description of how a vapor interface modifies the alpha relaxation process. In the context of ECNLE theory, this involves two distinct aspects. How does the interface and confinement modify: (i) the local hopping part of the problem as encoded in the dynamic free energy, and (ii) the long range displacement field and associated elastic barrier. Aspects (i) and (ii) are coupled, and formulating a theory for either one is a major challenge. 

Within the ECNLE theoretical framework, Mirigian and Schweizer \cite{38,39,40} constructed a zeroth order approach for free standing films based on the following two approximations which we explain here in the context of a one vapor interface thick film. For point (i), near the surface defined as $0 \leq z \leq r_{cage}$ (where for a sharp interface the center of particles of the first layer define $z=0$) caging constraints are softened due to losing nearest neighbors. The fraction of the bulk cage particles present at location z follows from geometry as \cite{38,39,40}:
\begin{eqnarray}
\gamma(z) = \frac{1}{2}-\left(\frac{z}{r_{cage}} \right)^3\left[\frac{1}{4}-\frac{3}{4}\left(\frac{r_{cage}}{z}\right)^2 \right].
\label{eq:9}
\end{eqnarray}
For $z = 0$, $\gamma(z) = 0.5$ corresponding to losing one half of the nearest neighbors. For $z = r_{cage}$, the full cage is recovered and $\gamma(z) = 1$. This is a highly local approximation, where surface-induced mobility is not propagated beyond the cage radius into the bulk. The dynamic free energy is modified as:
\begin{eqnarray}
F_{dyn}(r) = -3k_BT\ln\left(\frac{r}{d}\right) + \gamma(z)F_{caging}(r).
\label{eq:10}
\end{eqnarray}
Thus, near the surface all properties of the dynamic free energy (barrier, hopping time, localization length, local shear modulus) behave as a liquid with weaker dynamical constraints.

To address point (ii), the most naive (but no adjustable parameter) isotropic "cut off" assumption was adopted: the displacement field remains isotropic and of identical form as in the bulk, but it is set to zero at the vapor interface. Hence,  $u(r)$ inside the film is unchanged. Surely the basic cutoff idea is physically sensible, but the quantitative treatment is oversimplified. Within a cage radius of the surface, the single particle spring constant, $K_0$, of the dynamic free energy will be reduced. Hence, the elastic barrier is reduced, in a manner that depends on the location in the film (variable $z$) and determined by both the cut off effect and softening of $K_0=3k_BT/r_L^2$ within a cage radius of the interface. 

Prior studies \cite{38,39,40} using the above zeroth order extension of bulk ECNLE theory to free standing films have yielded sensible predictions for the temperature and film thickness dependence of the mobile layer, gradients of the shear modulus, local $T_g$ and alpha relaxation time, and film-averaged shifts of $T_g$. However, the naive cutoff approximation ignores all explicit anisotropy of the elastic fluctuations required for hopping and merits further investigation. In the next section we formulate what we think is a more realistic (but still approximate) treatment of the problem, and explore its consequences in subsequent sections. 

\section{Anisotropic Displacement Field with a New Boundary Condition}
\begin{figure}[htp]
\includegraphics[width=8cm]{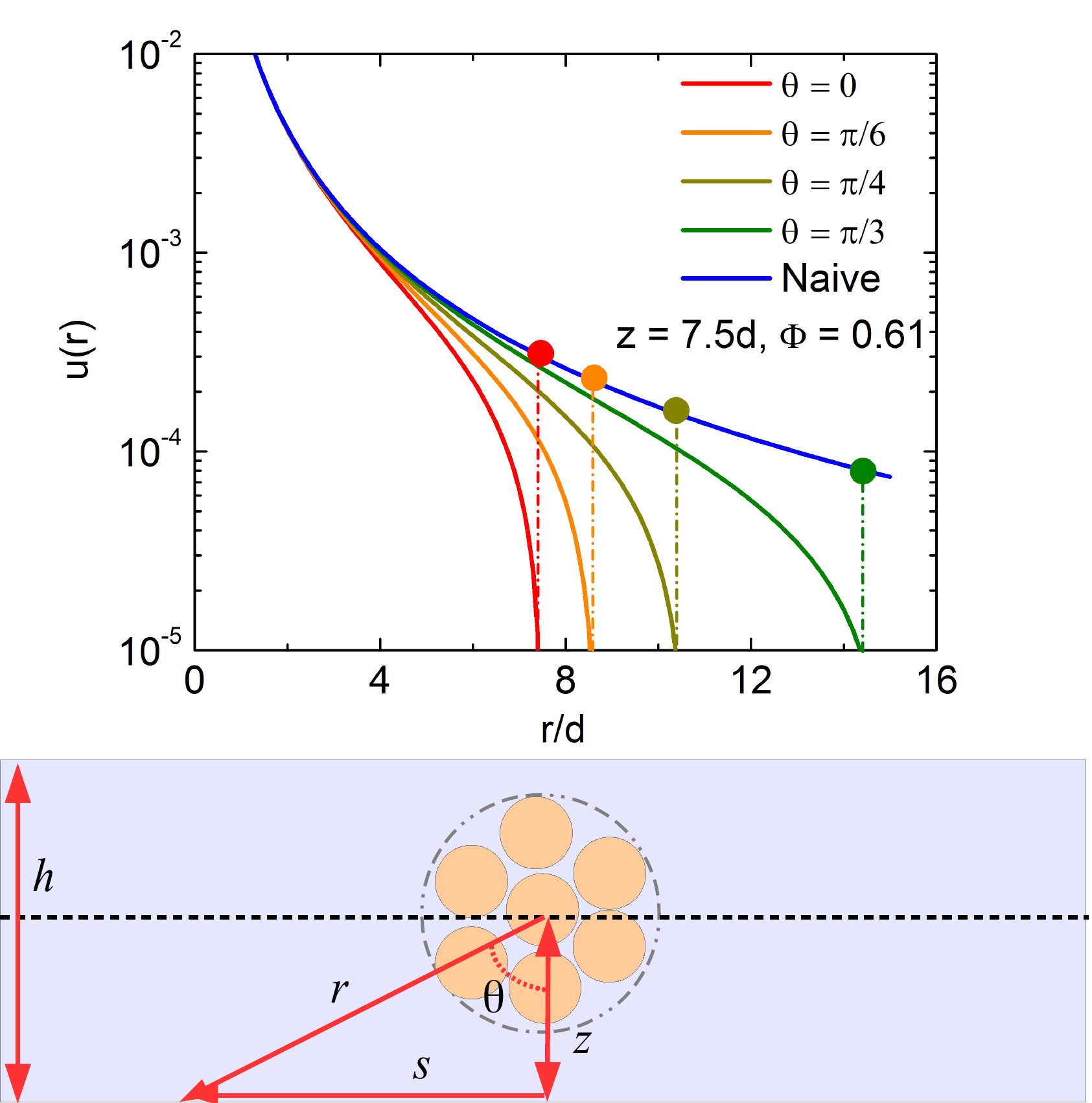}
\caption{\label{fig:2}(Color online) Anisotropic elastic displacement field (in units of particle diameter) in the lower half space as a function of distance $r$ from the cage center of a hopping event at $z = 7.5d$ as a function of the angle (defined in lower schematic) for a hard sphere fluid at $\Phi = 0.61$ (maps to $T_g$ of a thermal molecular liquid \cite{34}). The four points on the blue naive cut off model displacement field \cite{38} curve correspond to where the elastic field goes to zero based on the new anisotropic field model. }
\end{figure}

Rigorous solution of the problem of what the anisotropic facilitating collective elastic displacement field at the particle level is in thin films (generalization of Eq.(\ref{eq:2})) to allow local cage expansion is a complicated unsolved inverse problem. Here, given we model the liquid-vapor interface as perfectly sharp, we propose to improve our approach by requiring the displacement field continuously approaches zero at the vapor interface. Our prime motivation for doing so is to be internally consistent. That is, if the equilibrium density profile in the film is constant and precisely sharp, then allowing segments to dynamically penetrate the vapor region does not seem consistent. To allow analytic progress, we assume the displacement vector is controlled by the radial component $\mathbf{u}\equiv \mathbf{u_r}(r)$. This is an approximation which ignores the possibility that near the surface segment displacements parallel and perpendicular to the interface can be different. The asymmetric confinement of a thin film preserves azimuthal symmetry but $u(r)$ must vary with the angle $\theta$  (see Figure \ref{fig:2}) and the distance $z$ from the center of a hopping event to the interface. We assume the vapor layer modifies the displacement field in the lower half space (region I in Figure \ref{fig:1}), but for a thick film not in upper space (region II) which remains the same as in the bulk system. As a "trial displacement field" for $u(r)$ in region I, motivated as an anisotropic generalization of the general solution of Eq.(\ref{eq:2}), we adopt:
\begin{eqnarray}
u(r,\theta,z) = A_s(r,\theta,z)r+\frac{B_s(r,\theta,z)}{r^2},
\label{eq:11}
\end{eqnarray}
where $A_s(r,\theta,z)$ and $B_s(r,\theta,z)$ are chosen to enforce the boundary condition $u(r,\theta,z=0)=0$. The quantity $A_s$ must be zero in the bulk since the displacement field decays to zero at large distances, but is nonzero under finite confinement conditions. The field in the film exterior is zero.

For a hopping event with a cage centered at position $z$ in the film, we must enforce $u(r,\theta,z=0)=0$ and $u(r_{cage},\theta,z=0)=\Delta r_{eff}$. Straightforward analytic analysis gives in the lower space:
\begin{eqnarray} 
A_s(z,\theta)&=&-\frac{\Delta r_{eff}r_{cage}^2\cos^3\theta}{z^3-r_{cage}^3\cos^3\theta} = -\frac{\Delta r_{eff}r_{cage}^2}{(s^2+z^2)^{3/2}-r_{cage}^3}, \\
B_s(z,\theta)&=&\Delta r_{eff}r_{cage}^2\frac{z^3}{z^3-r_{cage}^3\cos^3\theta} \nonumber\\ &=& \Delta r_{eff}r_{cage}^2\frac{(s^2+z^2)^{3/2}}{(s^2+z^2)^{3/2}-r_{cage}^3}.
\label{eq:12}
\end{eqnarray}
where the variable "s" is defined in Figure \ref{fig:2}. Here $\Delta r_{eff}$ is its bulk value for $z \geq r_{cage}$ and smaller in a $z$-dependent manner based on the loss of neighbors effect. Note that for $z\rightarrow \infty$ or $\theta = \pi/2$, Eq.(\ref{eq:12}) reduces to $A_s=0$ and the bulk value of $B_s=\Delta r_{eff}r_{cage}^2$ is recovered. The elastic barrier follows as:
\begin{eqnarray} 
F_{elastic}(z) = \frac{3\Phi}{\pi d^3}\int_{V_film}d\mathbf{r}u^2(z,h,\theta,r)g(r)K_0(r,z).
\label{eq:13}
\end{eqnarray}

Figure \ref{fig:2} shows an example of the spatial variations of the new displacement field for a re-arrangement event centered at $z = 7.5d$. For the prior naive treatment \cite{38,39,40} it is direction ($\theta$)-independent. Enforcing $u(r,\theta,z=0)\rightarrow 0$ at the vapor interface introduces a stronger and anisotropic spatial dependence of the collective elastic field, and decreases its amplitude compared to the prior naive treatment. As $\theta$ decreases at fixed $z$, the displacement field more rapidly decays in space. For $z \leq 3.5d$, $u(r,\theta,z)$ is approximately independent of $\theta$ since it is dominated by the second term on the right hand side of Eqs.(12) and (13). At larger distances from the cage center, the influence of the confining term $A_sr$ on the displacement field becomes more significant.

\begin{figure}[htp]
\includegraphics[width=8cm]{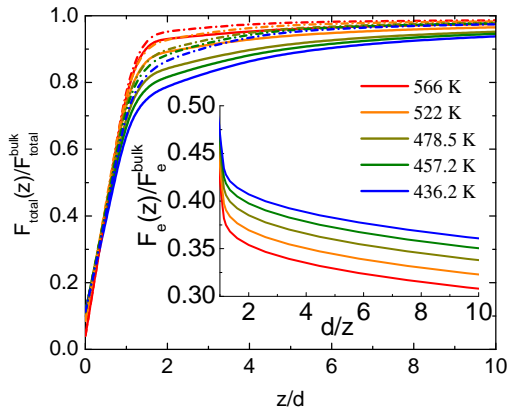}
\caption{\label{fig:3}(Color online) Total barrier (solid curves) normalized by its bulk value as a function of distance from the vapor surface of a thick film at several indicated temperatures for PS (bulk $T_g$ = 430K). The mapping corresponds to volume fractions of 0.55, 0.57, 0.60, and 0.61 for $T$ = 566, 522, 478.5, and 436.2 $K$, respectively. The dashed-dotted curves are the analogous results using the prior naïve displacement field cut off model \cite{38}. Inset: Normalized (to the bulk) collective elastic barrier plotted versus dimensionless inverse distance from the surface.}
\end{figure}

Figure \ref{fig:3} presents model calculations of the normalized total barrier gradient using the new and prior elastic displacement fields at different temperatures for PS. Since the new displacement field is significantly reduced compared to its bulk analog near the vapor interface, the total barrier estimated using the naïve displacement field recovers its bulk value on smaller distances. More generally, the total barrier becomes more suppressed near the interface and decays slower than its naïve analog. The variation of the barrier with location in the film has a "two-region" form due to the very different nature of the changes of the local (loss of nearest neighbors) and collective elastic (cutoff at surface) barriers upon confinement. The inset shows the new elastic barrier divided by its bulk value as a function of location in the film for various temperatures. The plot is made versus $d/z$ since we know (mentioned below Eq.(\ref{eq:5})) that in the bulk the elastic barrier slowly converges to its asymptotic bulk isotropic liquid value as the inverse distance from the cage center. The analog of this in the film is a long distance decay as $\sim d/z$.

\section{Thick Films: Gradients, Decoupling and Growing Mobile Layer Length Scale}
We apply the modified ECNLE theory to study spatial gradients of various properties for thick films. We then consider two new questions: (i) does z-dependent "dynamic decoupling" exist, and (ii) is there a connection between the bulk alpha time growth with cooling and a mobile layer length scale.
\subsection{Spatial Gradients}
\begin{figure}[htp]
\includegraphics[width=8cm]{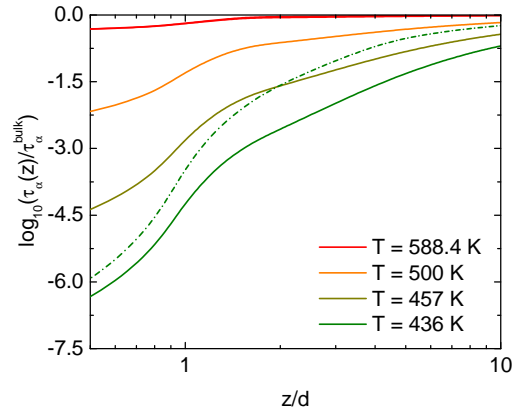}
\caption{\label{fig:4}(Color online) Log-log plot of the normalized (to the bulk) alpha relaxation time of a PS semi-infinite film as a function of distance from the surface at various temperatures. The dashed-dotted curve is the corresponding result for $T=436$ $K$ based on using the naïve displacement field cut off model \cite{38}.}
\end{figure}

Figure \ref{fig:4} shows calculations of the alpha relaxation time divided by its bulk value as a function of depth from the interface of a PS film at various temperatures. Relaxation massively speeds up near the vapor interface and enhanced mobility extends deep into the film. One also sees the new treatment of the elastic field results in a longer range gradient compared to the prior results. However, overall the differences between the new and prior alpha time gradient predictions are quantitative, not qualitative. The functional form of the alpha time gradient is generically different near and far from interface. This aspect appears to be different than seen in simulations \cite{14,46,47} performed at relatively high temperatures on atomic and simple polymer models which find a roughly "double exponential" variation, that is the logarithm of the alpha time varies exponentially with distance from the surface. Of course, simulations cannot address the deeply supercooled regime probed experimentally, and hence the validity of the double exponential relaxation time gradient in that regime is unknown. 

\begin{figure}[htp]
\includegraphics[width=8cm]{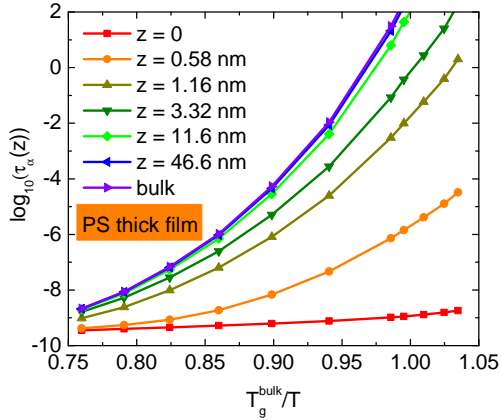}
\caption{\label{fig:5}(Color online) Angell-like plot of the alpha time (in seconds) versus inverse normalized temperature at various indicated distances z from the free surface of a semi-infinite PS film.}
\end{figure}

Figure \ref{fig:5} presents Angell-plot type calculations of the PS film alpha relaxation time versus normalized inverse temperature over a wide range of locations in the film. The rate of growth of the alpha time with cooling varies enormously as one moves from near the surface into the film interior. The temperature dependence is extremely weak very close to the surface. By eye on the scale of the graph, bulk-like behavior is attained at $\sim 15-20$ nm from the surface. 

\begin{figure}[htp]
\includegraphics[width=8cm]{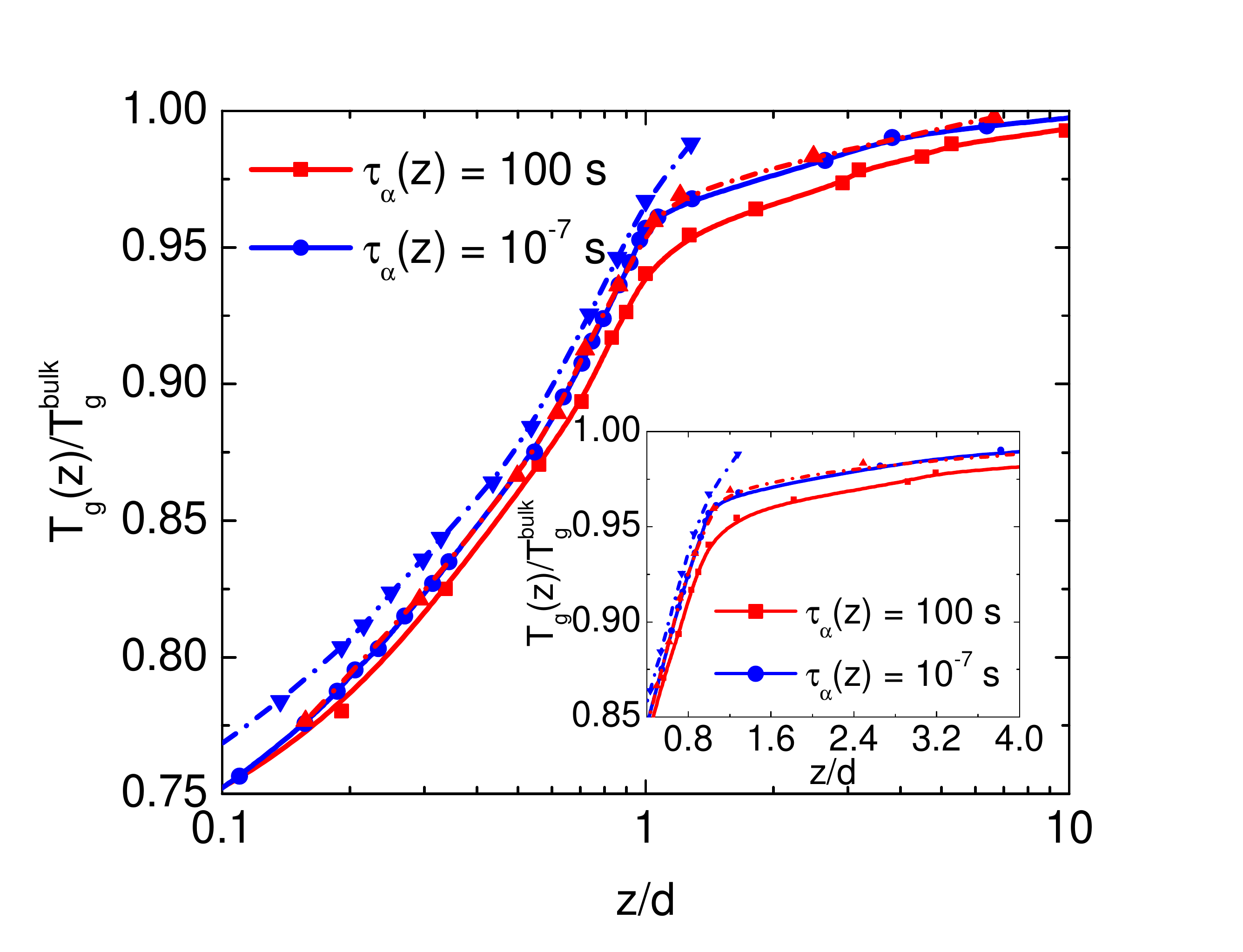}
\caption{\label{fig:6}(Color online) Linear-log plot of the local glass transition temperature normalized by its bulk value as a function of location in the semi-infinite PS film based on two different vitrification criteria, 100 s and 100 ns. The dashed-dotted curves corresponds to the analogous results using the naive displacement field cut off model \cite{38}. Inset: same results as in the main frame plotted in a linear-linear format.}
\end{figure}

From knowledge of how the alpha time grows with cooling as a function of location in the film, the spatial gradient of $T_g$ can be calculated. Results are shown in Figure \ref{fig:6} in a vertically normalized log-linear format. A very rapid variation is predicted within a cage radius of the interface due to the loss of neighbors effect, followed by a slow drift towards the bulk value at larger distances from the interface where collective elasticity effects dominate the thin film perturbation due to the cut off of the displacement field at the surface effect. Such a two-regime variation in space of the local $T_g$ may perhaps be viewed as providing modest support for the qualitative suggestion \cite{23} that large film-averaged reductions of $T_g$ are mainly a "surface effect". 

Results are also shown in Figure \ref{fig:6} using a much smaller vitrification timescale criterion of only 100 ns, motivated mainly by simulation studies. Of course, the bulk $T_g$ also changes (increases) when the vitrification criterion is changed, which is taken into account. In the plotting format, we surprisingly find that using a vitrification criterion corresponding to 9 decades faster dynamics yields normalized $T_g$ gradients almost identical to those obtained with the typical experimental criterion (100 s), at least close to the interface. Farther from the interface there are noticeable, but still not dramatic, differences. The latter are expected since the relative importance of local versus elastic effects is very different at temperatures where the bulk alpha time is 100 s versus 100 ns. We also show results based on using the prior naive displacement field. The differences compared to the new approach are modest, but the prior approach predicts smaller $T_g$ reductions that extend less into the film.  

\subsection{Mobile Layers and Correlation with the Bulk Alpha Relaxation}
From our results for the alpha time gradient, we define a mobile layer thickness based on a chosen criterion. Following recent simulation studies \cite{29,31,47} we adopt the criterion of a mobile layer thickness, $\xi$, as the distance from the film surface where the relaxation time reaches a fraction $C$ of its bulk value at a fixed temperature. The results of this calculation for PS are shown for $C=0.5$ and 0.8 in the inset of Figure 7. As must be true, $\xi(T)$ grows with cooling, and it does so quite strongly. The reason is that as temperature decreases, the collective elastic contribution to the barrier becomes more important relative to the local cage barrier, and it is more sensitive to the vapor interface via the elastic field cut off effect. For example, $\xi\approx 25d\approx 30$ nm near the bulk PS $T_g$ for a criterion of $C=0.5$. This large absolute length scale again reflects the importance of long range elastic effects in determining the mobility gradient far from the surface.

Motivated by simulation \cite{29,31,46,47} general theoretical considerations \cite{10} and our recent study in the bulk \cite{48} the main frame of Figure 7 explores a possible exponential correlation between the bulk alpha relaxation time and the temperature-dependent mobile layer length scale. Remarkably, over 15 decades we find the logarithm of the bulk alpha relaxation time increases in a weakly sub-linear manner described (essentially equally well) by the following two forms:
\begin{eqnarray} 
\ln(\tau_{\alpha,bulk})  \varpropto \left(\xi(T)/T \right)^{3/4} \varpropto \left(\xi(T)/d \right)^{0.8}.
\label{eq:15}
\end{eqnarray}
This behavior is identical to what we recently found for isotropic liquids based on a growing bulk cooperativity length scale defined as when the alpha time achieves its nearly full value \cite{48}. This behavior also applies well if one only analyzes our calculations over the typical simulation time scale range ($\sim$ 1 ps to 10-100 ns) where the collective elastic effects are quite small (not shown).

\begin{figure}[htp]
\includegraphics[width=8cm]{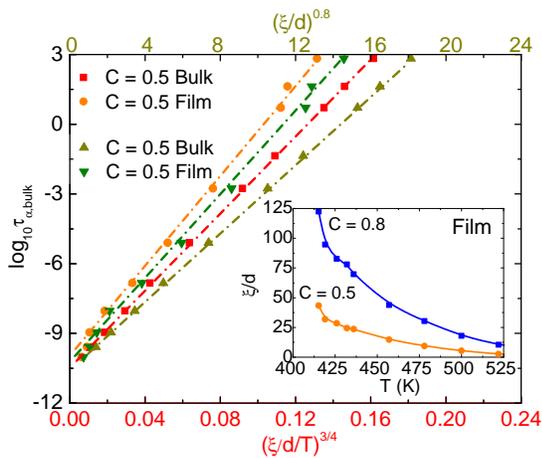}
\caption{\label{fig:7}(Color online) Logarithm of the bulk PS liquid alpha relaxation time (in secs) as a function of $\left(\xi(T)/d/T \right)^{3/4}$ (red and orange curves) and $\left(\xi(T)/d \right)^{0.8}$ (green and dark yellow curves), where $\xi$ is the mobile layer thickness in the free-standing thick film computed based on the criterion $\tau_\alpha(\xi)/\tau_\alpha^{bulk}=0.5$. Inset: the temperature-dependent mobile layer length scale for two specific numerical criteria ($C$ values).}
\end{figure}
\subsection{Interfacial Dynamic Decoupling}
In bulk glass-forming molecular liquids, due to space-time dynamic heterogeneity the phenomenon of "decoupling" is observed in the deeply supercooled regime \cite{10,49}. This corresponds to the temperature dependence of the diffusion constant becoming weaker than that of the viscosity or single molecule reorientation time. The physics of this effect is still not well understood, but such dynamic heterogeneity is presumably also present in thin films and is not included in our present theoretical approach. In glass-forming polymer melts, there is another type of decoupling in the deeply supercooled regime where, to strongly varying degrees depending on fragility, chain scale relaxation times grow more weakly with cooling than the segmental relaxation time \cite{50,51}. This phenomena is sometimes described by an effective power law relation:

\begin{eqnarray} 
\frac{\tau_{chain}(T)}{\tau_\alpha(T)} = \tau_\alpha(T)^{-\Delta},
\label{eq:16}
\end{eqnarray}
where the decoupling exponent $\Delta$ varies from $\sim$ 0.5 to nearly zero depending on chemistry \cite{51}.

On the other hand, in films there is explicit spatial heterogeneity of relaxation and mass transport due to the presence of interfaces. This raises the question, first clearly expressed by Simmons \cite{52}, of whether a film location dependent decoupling behavior might apply in the sense that:
\begin{eqnarray} 
\frac{\tau_\alpha(z,T)}{\tau_{\alpha,bulk}(T)} \varpropto \tau_{\alpha,bulk}(T)^{-\varepsilon(z)},
\label{eq:17}
\end{eqnarray}
where here the "decoupling exponent" varies with location in the film. If applicable, Eq.(\ref{eq:17}) implies a remarkable factorization of the effective barrier into its temperature and chemistry dependent bulk value multiplied by a purely z-dependent function:
\begin{eqnarray} 
F_{total}(z,T) \approx F_{total,bulk}\left(1-\varepsilon(z) \right).
\label{eq:18}
\end{eqnarray}
Why such a factorization could be true is not at all obvious to us. It surely is not a trivial generic consequence of the existence of a large mobility gradient. Since experiments have not been able to measure with fine spatial resolution the alpha relaxation time as a function of location in a film over a wide range of temperatures, direct experimental testing of such a decoupling phenomenon does not seem feasible at present. However, simulations can examine this question, at least in the lightly supercooled regime they can be performed. Moreover, if such decoupling exists it should have some distinctive consequences for the more averaged quantities that are experimentally measurable, although discussion of this point is beyond the scope of the present article.

In the context of ECNLE theory, given there are two barriers that determine the alpha relaxation time, with different temperatures dependences and different variations with location in the film, it is a priori very unclear whether Eq.(\ref{eq:18}) holds. One possible reason why it might hold is the foundational idea of ECNLE theory of confined films that a dynamic free energy can be locally constructed at each location in a film which determines all physical quantities needed to compute a hopping time.

\begin{figure}[htp]
\includegraphics[width=8cm]{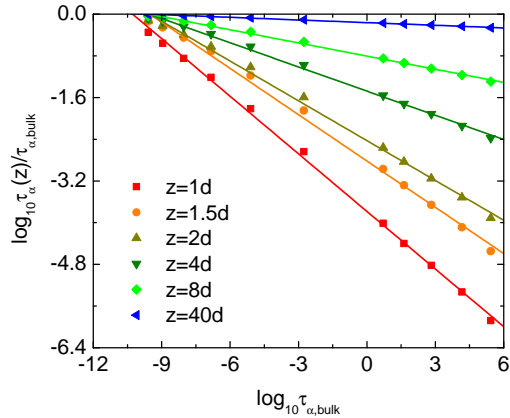}
\caption{\label{fig:8}(Color online) Double logarithm (base 10) plot of $\tau_\alpha(z)/\tau_{\alpha,bulk}$ as a function of the bulk PS alpha relaxation time at various distances from the vapor surface of a thick film. The lines are power law fits through all the theoretical data points which span 15 decades in bulk alpha time.}
\end{figure}

Figure \ref{fig:8} presents representative calculations that explore the above possibility for PS over an exceptionally wide range of temperature corresponding to the bulk alpha time varying by 15 decades. Rather remarkably, the almost perfect straight lines in the log-log representation show decoupling is predicted. If one analyzes only the shorter time scale regime relevant to simulations, the apparent slopes are a bit smaller than if one fits the theoretical data to a single power law over 15 decades, but the qualitative behavior is the same. The existence of this decoupling behavior, and hence the validity of Eq.(\ref{eq:18}), is in qualitative accord with the very recent simulation discovery of such decoupling by Simmons and coworkers \cite{52}.

\begin{figure}[htp]
\includegraphics[width=8cm]{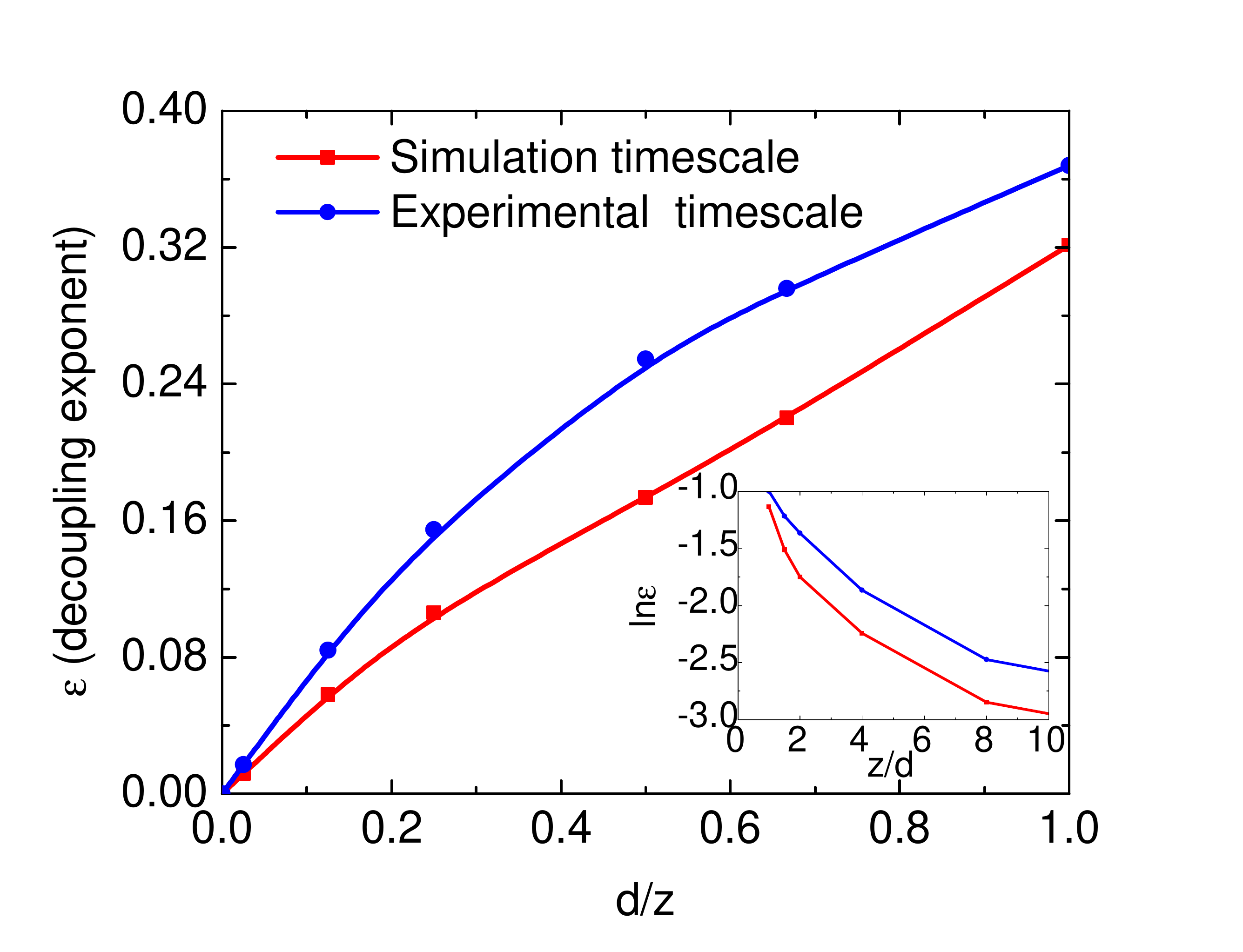}
\caption{\label{fig:9}(Color online) Decoupling exponent determined from Figure \ref{fig:8} as a function of dimensionless inverse distance from the surface.  "Experimental time scale" curve indicates all of the data in Figure \ref{fig:8} was used, while "simulation timescale" curve only fits the first 5 decades of data in Figure \ref{fig:8} to extract an exponent. Inset: Natural log-linear plot of the decoupling exponent versus location in the thick film.}
\end{figure}

The corresponding decoupling exponents extracted from Figure \ref{fig:8} using the entire range of data (covers the experimental timescale, blue curve) and also just the first 5 decades (covers typical simulation timescale, red curve) are plotted versus the inverse distance from the surface in the main frame of Figure \ref{fig:9}. The decoupling exponent is larger if the deeply supercooled regime results are included, as physically expected. Both calculations reveal that, very roughly, $\varepsilon$ increases linearly with $1/z$ at large $z$. This variation is a direct consequence of the dominance of the elastic field cut off effect far from the interface, which can be shown to provide a correction to the dynamic barrier that scales inversely with $z$. The latter follows by ignoring $F_B$ in Eq.(\ref{eq:7}) and assuming $K_0(z)\sim K_{0,bulk}$, from which one can analytically derive an approximate inverse in distance variation from the surface behavior 
\begin{eqnarray} 
\log_{10}\frac{\tau_\alpha(z)}{\tau_{\alpha,bulk}} \varpropto -\frac{r_{cage}}{4z}\log_{10}\tau_{\alpha,bulk}.
\label{eq:19}
\end{eqnarray}

However, closer to the surface (smaller $z$) there are strong deviation from an inverse power law behavior since here reduction of the local barrier via the loss of cage neighbors effect becomes very important and varies with film location in a very different manner than $z^{-1}$.

Overall, the predicted decoupling-related trends appear to be similar to recent simulations in the dynamic crossover regime \cite{47,52} except the latter find an exponential decay of the decoupling exponent. The inset of Figure \ref{fig:9} shows we do not predict such an exponential behavior. This deviation is likely related to the fact we do not predict a roughly double exponential variation of the alpha time with location $z$ (see Figure \ref{fig:4}), suggesting missing physics in the theory and our treatment of points (i) and (ii) stated in section III. 
\section{Thin Films}
We now study free-standing films of thickness, $h$. This system has been analyzed in prior ECNLE theory articles \cite{38,39,40} based on the naive cut off treatment of the elastic field. Our goals here are to first contrast the predictions of the prior and new anisotropic displacement field treatment for the barrier gradient. We then use the improved version of the theory to perform new quantitative studies of $T_g$ shifts as a function of film thickness, vitrification criterion and polymer fragility. No adjustable parameter comparisons are made with experiment and simulation.
\subsection{Position-Dependent Barriers}
\begin{figure}[htp]
\includegraphics[width=8cm]{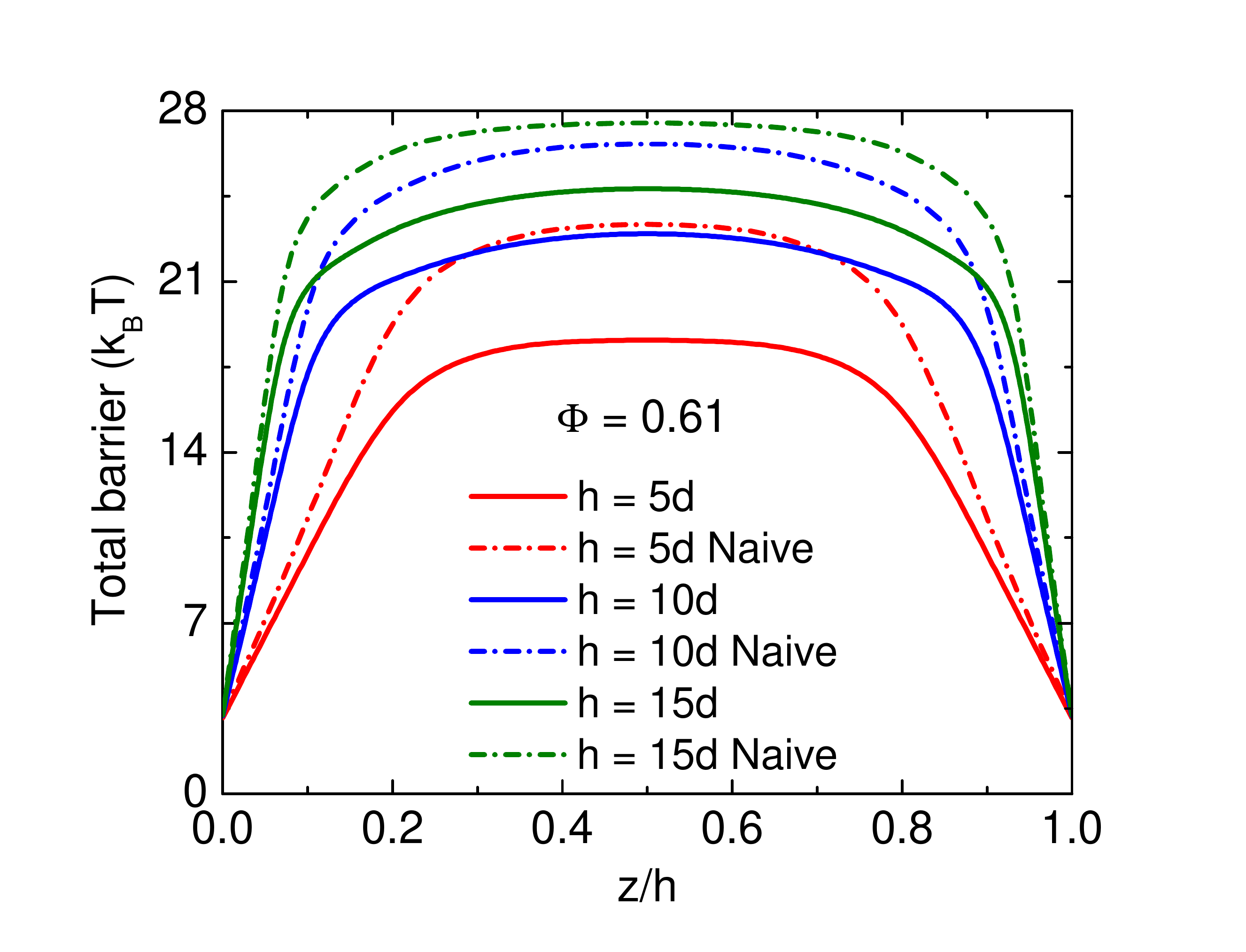}
\caption{\label{fig:10}(Color online) Total barrier as a function of location in thin films of thickness 5, 10 and 15 $d$ at a high volume fraction corresponding to $T_g$ of the bulk PS system. For all locations the barrier is less than its value in the bulk. Results are shown based on the new (solid) and old naive \cite{38} (dashed-dot) treatments of the elastic displacement field.}
\end{figure}

Figure \ref{fig:10} shows calculations of the total barrier gradient for the underpinning hard sphere fluid at $\Phi = 0.61$ (maps to the bulk $T_g$ for molecular liquids \cite{34} where the barrier is $\sim$ 32 $k_BT$) for three values of film thickness as a function of normalized location in the film. One sees huge reductions of the barrier at/near the surface. Moreover, the reductions are nearly independent of film thickness and almost identical for the two treatments of the elastic field since the physics is on the local cage scale (loss of neighbors effect). Deeper into the film, the barrier initially grows in a roughly linearly manner, and is smaller based on the new treatment of the elastic field. The different slopes for different thickness films are a trivial consequence of the normalization of $z$ by film thickness, $h$. The barrier tends to saturate in the middle of the film, but below its bulk value, and decreases more as the film thins (more elastic barrier suppression and coupling between effects at the two interfaces) and for the anisotropic treatment of the displacement field. Overall, these large changes in barriers immediately translate to orders of magnitude speed up of the alpha relaxation time.

\subsection{Thickness-Dependent Film-Averaged Alpha Relaxation Times, $T_g$  shifts and Influence of Vitrification Criteria}

\begin{figure}[htp]
\includegraphics[width=8cm]{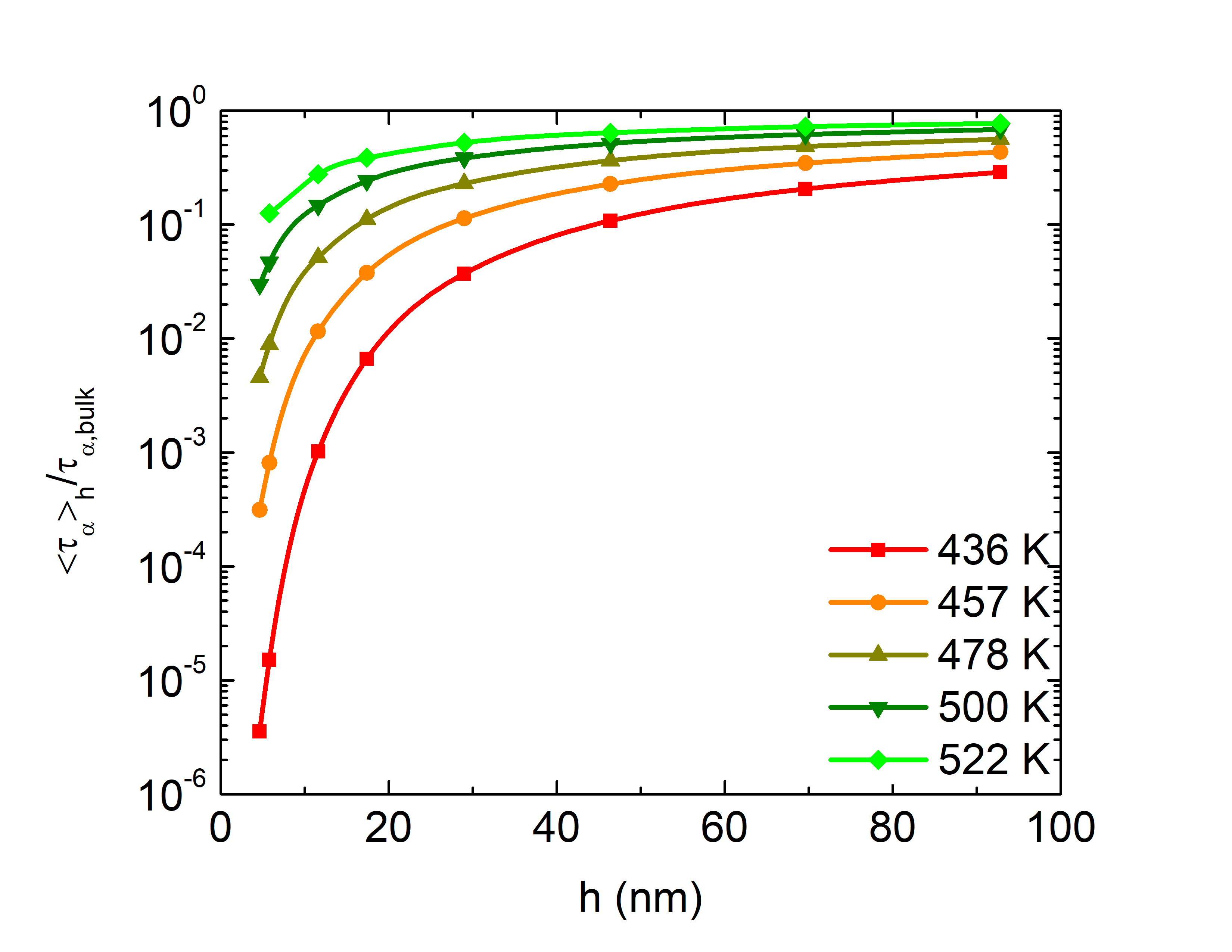}
\caption{\label{fig:11}(Color online) Normalized film averaged relaxation time as a function of the film thickness at several temperatures from 522 $K$ to 436 $K$ for PS (bulk $T_g$ = 430 $K$).}
\end{figure}

Figure \ref{fig:11} presents results for the film-averaged alpha relaxation time normalized by its bulk value as a function of film thickness at five temperature for the PS system. Very crudely, the basic shape of the curves is akin to the alpha time gradient plots of Figure \ref{fig:4}. However, given the spatial gradient is averaged over, all curves display a much reduced "two-regime" form than seen in Figure \ref{fig:4}. Nevertheless, the slowly decaying nature of the curves at large values of h is still evident. For ultra-thin 4 nm films, the average alpha time is predicted to be 1 to $\sim$ 5.5 orders of magnitude faster than the corresponding bulk value as the film is cooled from 522 $K$ to 436 $K$. With decreasing temperature, the film thickness required to recover the bulk alpha time also grows quite strongly, reflecting the increasing relative importance of the long range collective elastic barrier. 

We now consider film-averaged $T_g$ shifts. As discussed in the literature \cite{1,2,53} and our prior papers \cite{38,39,40} there are two main approaches to determine the thickness-dependent  , "pseudo-thermodynamic" and "dynamic". We employ our alpha time gradient calculations to study the thickness-dependent pseudo-thermodynamic $T_g$ based on a computed local $T_g(z)$ defined by when reaches a specified vitrification criterion: 
\begin{eqnarray} 
\langle T_g \rangle _h = \frac{1}{h}\int_0^h T_g^{local}(z)dz.
\label{eq:20}
\end{eqnarray}

This $T_g$ is thought to be probed in thermodynamic measurements such as heat capacity and ellipsometry. We also compute the corresponding dynamic glass transition temperature as when the film-averaged alpha time reaches the chosen vitrification criterion (specified by the exponent $y$ below):
\begin{eqnarray} 
\langle \tau_\alpha(T_g) \rangle _h = \frac{1}{h}\int_0^h \tau_\alpha(z;T_g) dz =10^y \ce{s}.
\label{eq:21}
\end{eqnarray}
This $T_g$ value is believed to be relevant to dynamic measurements such as dielectric spectroscopy.

\begin{figure}[htp]
\includegraphics[width=8cm]{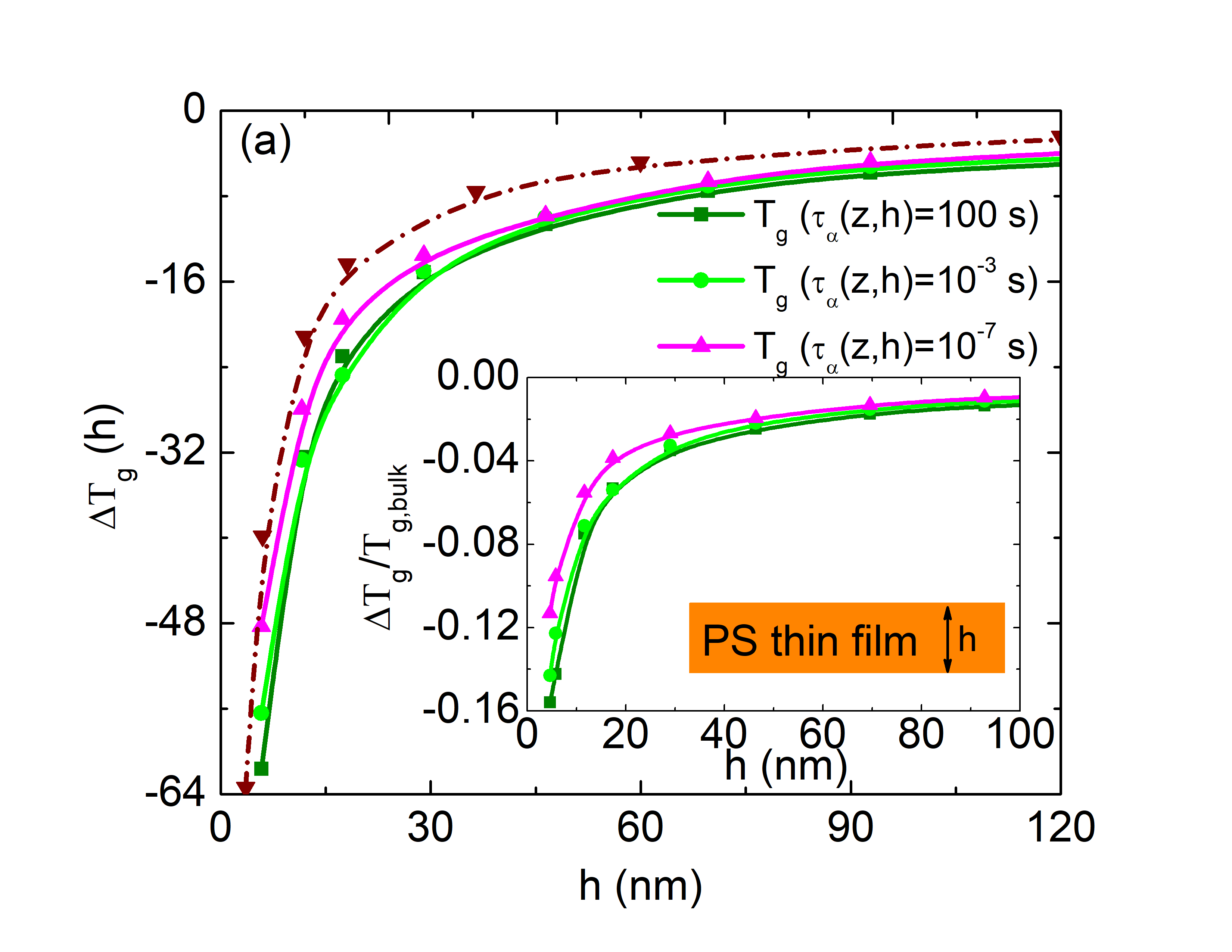}\\
\includegraphics[width=8cm]{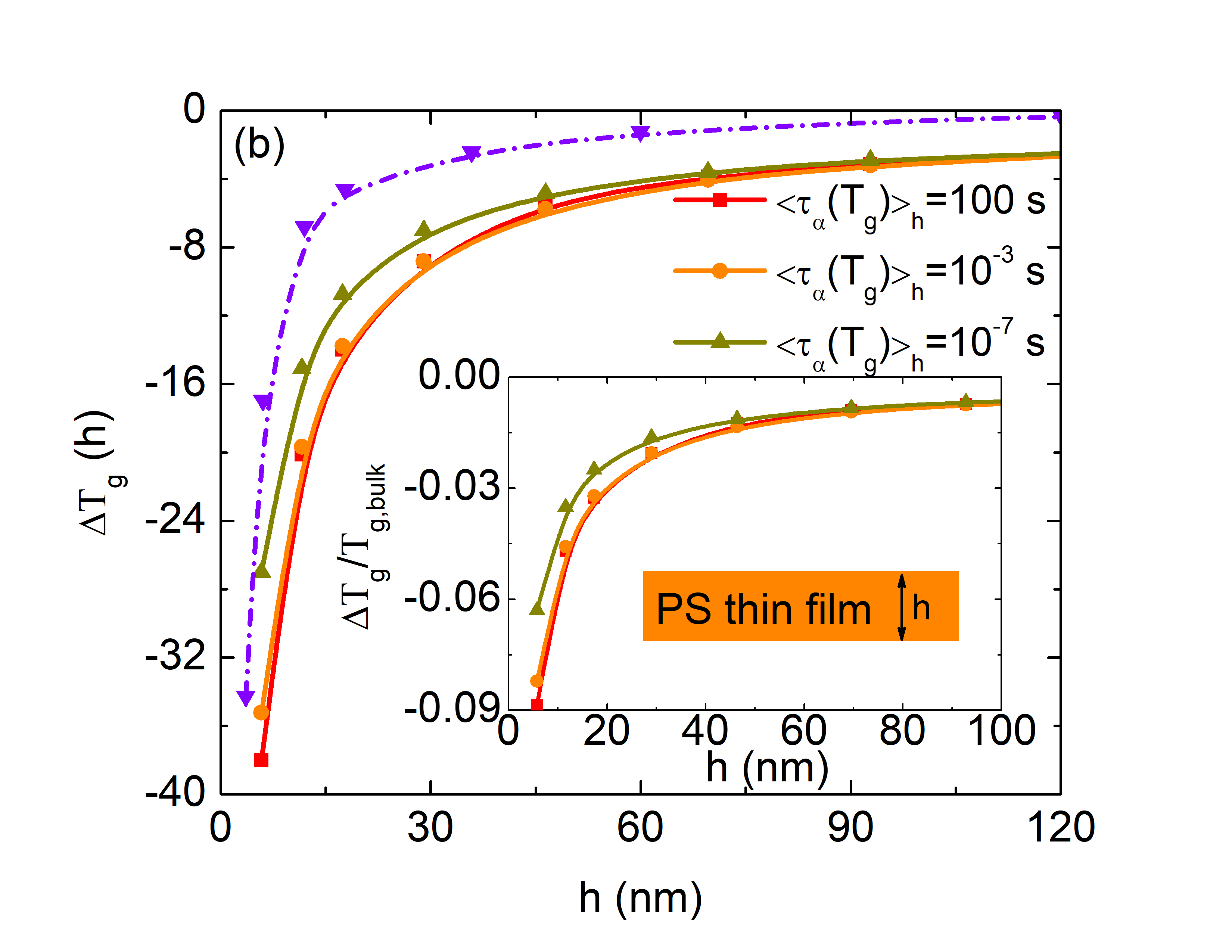}
\caption{\label{fig:12}(Color online) Film-averaged glass transition temperature shift of PS films (in Kelvin) as a function of film thickness (in nm) for two different protocals of extracting a film-averaged glass transition temperature and 3 vitrification timescale criteria. Pseudo-thermodynamic calculations (a) based on $\left< T_g \right>_h$  are shown for vitrification criteria of $\tau_\alpha(z,h)=100$, $10^{-3}$ and $10^{-7}$ s as indicated by the dark green, light green, and pink lines, respectively. Analogous dynamic calculations (b) of $T_g(h)$ based on the vitrification criteria $\left< \tau_\alpha(T_g) \right>_h = 100$, $10^{-3}$ and $10^{-7}$ s correspond to the red, orange, and yellow green lines, respectively. The dashed-dotted brown and violet curves are the corresponding pseudo-thermodynamic and dynamic calculations for a vitrification criterion of 100 sec per previous work based on the naive prior treatment of the elastic field \cite{38,39,40}. Inset: normalized film-averaged glass transition temperature depression of PS films determined using the pseudo-thermodynamic approach for three vitrification time-scale criteria plotted versus the film thickness in nanometers.}
\end{figure}

Figure \ref{fig:12} presents calculations of the $T_g$ depression (in Kelvin) for PS as a function of the film thickness (in nm) for the two different definitions of the glass transition and three different vitrification criteria that span 9 decades. We note that our motivation for exploring the role of vitrification time scale criterion is mainly theoretical, and is also relevant to simulations which effectively adopt a time scale for defining $T_g$ that is many orders of magnitude shorter than experiment. 

The pseudo-thermodynamic results in Figure \ref{fig:12}a show a substantially larger drop than the dynamic analog of Figure \ref{fig:12}b as a consequence of different mobility gradient averaging. Although the naive elastic field model approach (dashed-dot curves) exhibits the same trends as the new treatment, their $T_g$ reductions differ rather significantly, e.g., by nearly 20 $K$ for $h\sim 5$ nm. Reducing the vitrification time scale criterion leads to a smaller $T_g$ depression (long range elastic barrier less important), although the overall shape of the curves are not very different. The inset of Figure \ref{fig:12}a re-plots the main frame pseudo-thermodynamic results by normalizing temperature shifts by $T_{g,bulk}$. The corresponding $\Delta T_g/T_{g,bulk}$ curve is nearly identical for the 100 s and 0.001 s vitrification criteria. On the other hand, for a 100 ns criterion one sees non-negligible deviations in the direction of smaller $T_g$ suppression. The trends in the inset of Figure \ref{fig:12}b are qualitatively the same as those in Figure \ref{fig:12}a. Overall, the modest sensitivity of $T_g$ shifts plotted in the two formats to the vitrification criterion adopted would appear to be good news for the relevance of simulations to experimental behavior for this specific question. However, its implications for understanding the puzzling "cooling rate $T_g$ measurements" of Fakhraai \emph{et al} \cite{25} is unclear.
\subsection{Comparison to Experiment and Simulation}
\begin{figure}[htp]
\includegraphics[width=8cm]{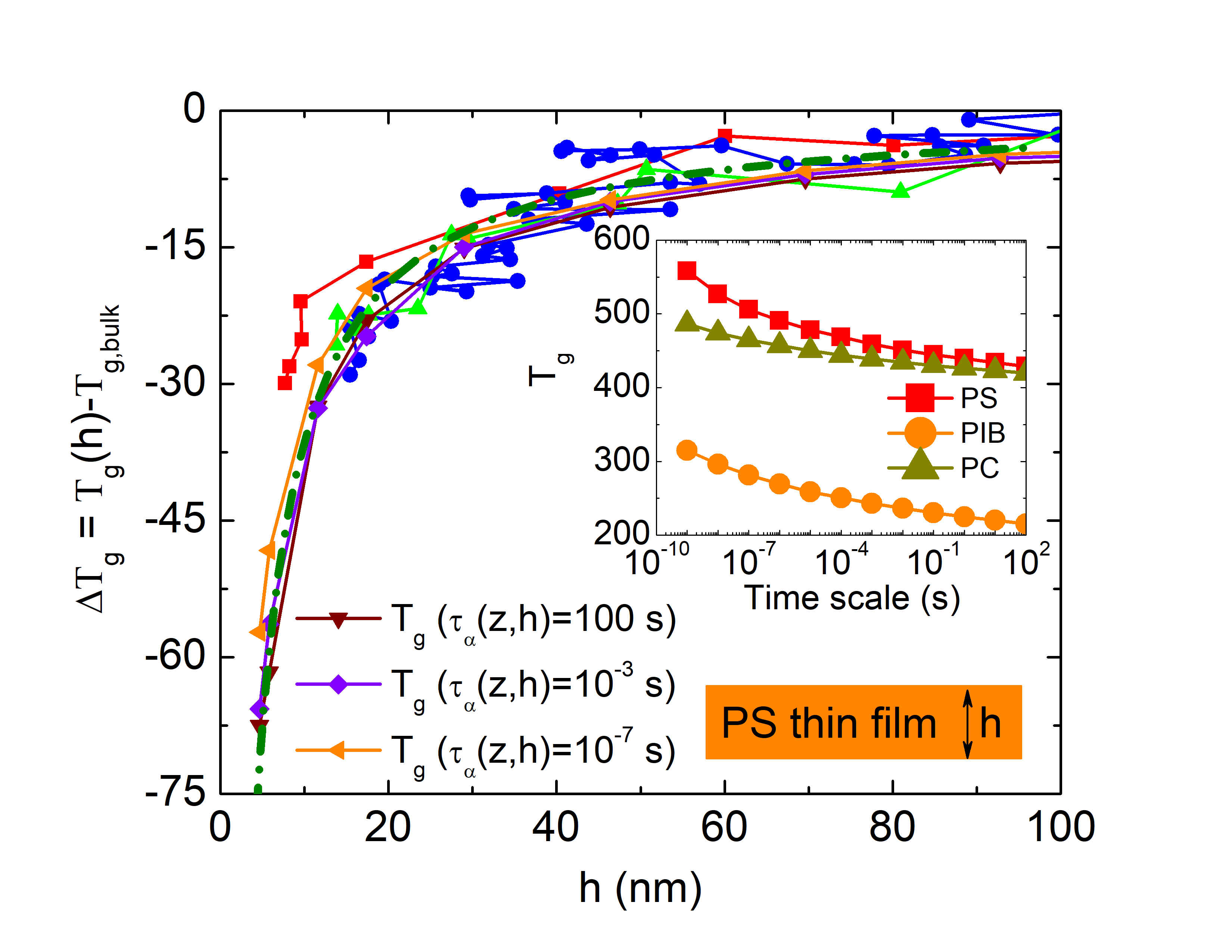}
\caption{\label{fig:13}(Color online) Film-averaged glass transition temperature shift of PS thin films determined by the pseudo-thermodynamic approach for three vitrification criteria as a function of the film thickness in nm (brown, purple and orange data points with interpolating curves drawn through them). The red, blue, and green curves correspond to a simulation result \cite{55}  and experimental data of refs \cite{54} and \cite{19}, respectively. The green dashed-dotted curve (barely visible) is a fit using Eq.(\ref{eq:22}). Inset: glass transition temperature of bulk PS, PIB and PC polymer melts as a function of vitrification time scale criteria.}
\end{figure}

Figure \ref{fig:13} presents quantitative no adjustable parameter calculations of the pseudo-thermodynamic $T_g$ shifts of PS as a function of film thickness for three vitrification criteria. Also shown are two sets of experimental data \cite{19,54} and one set of simulation data \cite{55} using a lightly coarse grained model of polystyrene. The two experimental data sets largely overlap in a statistical sense to within the significant experimental uncertainties. Given the latter, our goal is to compare the theory results with experiment in a global, not individual, manner. The simulation results employed a definition of the glass transition as when the alpha time is only 1 ns. Since the bulk $T_g$ in the theory calculations varies with vitrification criterion, the inset shows ECNLE theory results for the three polymers studied.

Overall, we think Figure \ref{fig:13} shows good agreement between theory, experiment and simulation. The experimental data extends down to $h \sim 12-15$ nm, where $T_g$ is reduced by $\sim$30 $K$ for PS. To achieve such a suppression in simulation requires a much thinner film of $\sim$7 nm. This is perhaps expected given the simulation vitrification criterion corresponds to a much shorter relaxation time, a trend also consistent with our theoretical calculations.

An empirical analytical expression used by many to fit experiments and simulations of thickness-dependent $T_g$ shifts motivated by a "2-layer" model of film dynamics is \cite{14,38,56}:
\begin{eqnarray} 
T_g(h) = \frac{T_{g,bulk}}{1+\Gamma/h},
\label{eq:22}
\end{eqnarray}
where $\Gamma$ is an adjustable fit parameter. Fits of Eq.(\ref{eq:22}) to our theory calculations (green dashed-dot curve in Figure 13) using the 100 s time criterion reveal good agreement with $\Gamma\approx 0.92d$. However, we emphasize that this comparison has been made only because some experimentalists continue to fit their data to such a two-layer model as an empirical exercise. We are not advocating the veracity of such a naive model. Indeed, our theory predicts a continuous gradient of alpha times and local $T_g$'s, and the ability of Eq.(\ref{eq:22}) to fit our theoretical data does not provide evidence a 2-layer model is correct.
\subsection{Influence of Polymer Fragility}
\begin{figure}[htp]
\includegraphics[width=8cm]{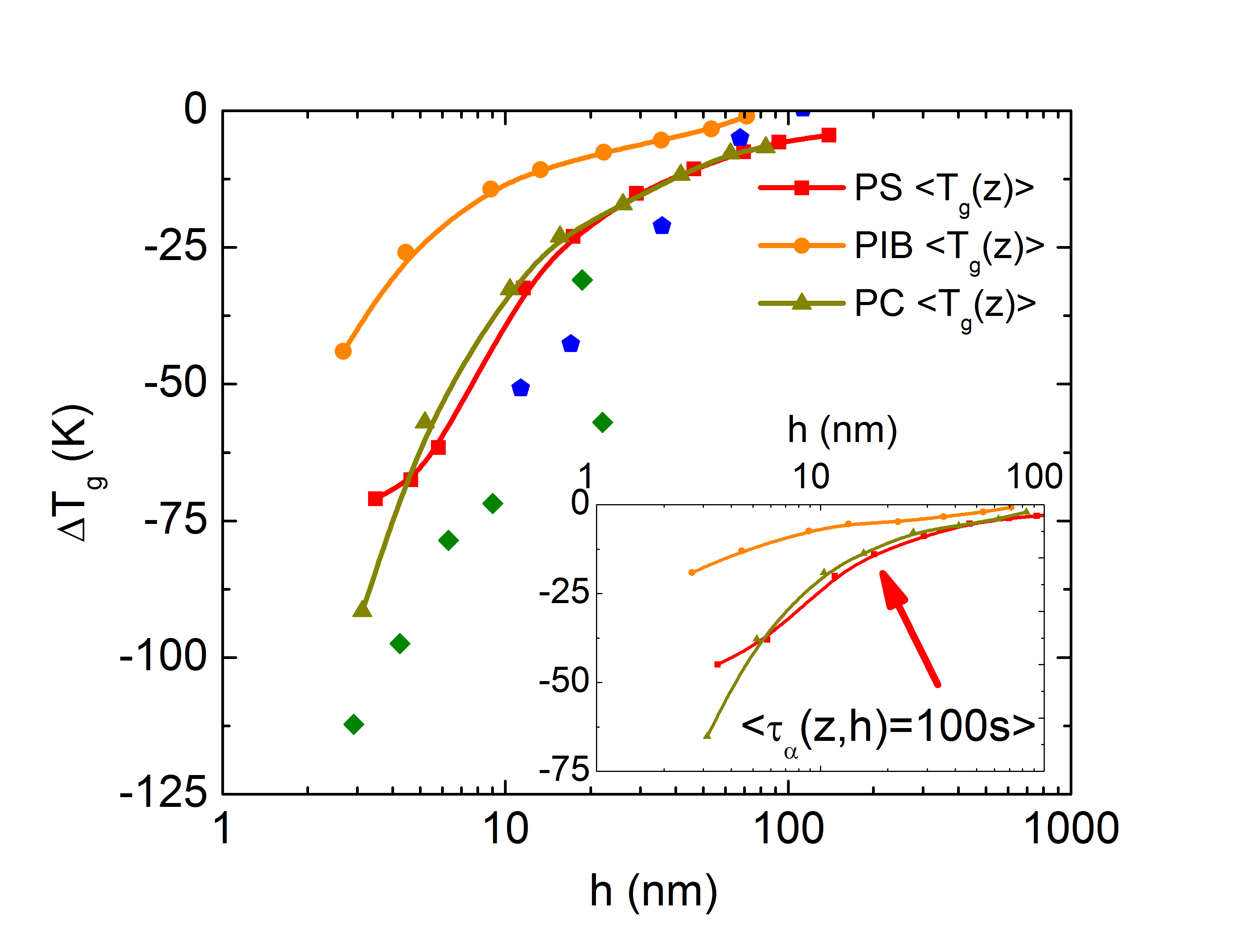}
\caption{\label{fig:14}(Color online) Film-averaged glass transition temperature shift for PS, PC and PIB calculated using the pseudo-thermodynamic (main frame) and dynamic (inset) approach as a function of film thickness in nm. The $\left< T_g\right>_h$ calculation uses a local vitrification criterion of $\tau_\alpha(z,h)=100$; the dynamic calculation of $T_g$ uses the global criterion of $\left< \tau_\alpha(T_g)\right>_h =100$ s. Green triangle and diamond points corresponding to experimental data of ref \cite{57} and ref \cite{58} for PC and PS thin films, respectively.}
\end{figure}

Figure \ref{fig:14} shows the influence of polymer fragility (and hence variable relative importance of collective elastic versus local cage barriers) on thin film pseudo-thermodynamic (main frame) and dynamic (inset) $T_g$ shifts based on the 100 s local vitrification criterion using the new anisotropic elastic field approach. Results are presented for PC, PS, PIB (bulk theoretical fragilities from ref \cite{37} are $\sim$ 140, 110, 46, respectively). Also shown as discrete experimental data points in the main frame for PS \cite{57} and PC \cite{58} based on the bubble inflation creep method of McKenna et al \cite{59}.

The absolute $T_g$ suppression in degrees Kelvin is much smaller for PIB, but similar for PS and PC, for both methods of determining a film-averaged glass transition temperature. Consider first the results in the main frame of Figure \ref{fig:14}. The experimental data display modestly larger shifts than our theoretical calculations. The remarkable reduction of $\sim 100$ $K$ for a PC film of $\sim$ 4 nm observed experimentally is captured quite well. The large film thickness required for the $T_g$ shift to disappear also seems well captured by the theory for PS, and we predict this length scale should be very similar for PC. However, there must be a nonuniversal aspect to the question of the film thickness required to eliminate confinement effects. First, if film thickness is expressed in absolute units (e.g., nm), then the elementary microscopic length scale (e.g., Kuhn segment diameter in our model) enters which is polymer specific; for example, it is almost a factor of 2 smaller for PIB compared to PS or PC. But even if film thickness is non-dimensionalized by this local chemical length scale, we expect the reduced film thickness to recover bulk behavior will increase with fragility. The reason is that higher fragility in the bulk liquid ECNLE theory arises from a greater relative importance of the collective elastic component of the dynamic activation barrier \cite{37}, which in turn is more sensitive to film thickness (via the cut off effect) than the local cage scale component of the barrier. This aspect is clear from Figure \ref{fig:14} where one sees that the low fragility PIB system recovers its bulk $T_g$ at a significantly smaller film thickness than PS or PC.

Overall, given the noise in the experimental data, the approximate nature of the theory, the lack of any fitting parameters, and the fact we do not compute the precise observable measured experimentally, we find the agreement between theory and experiment encouraging. The inset of Figure \ref{fig:14} shows the analogous theoretical results based on the dynamic definition of the film-averaged $T_g$. All polymer chemistry trends are unchanged, but the maximum $T_g$ suppressions are smaller by a factor of $\sim 1.5-3$. 
\subsection{Caveats}
Finally, we mention some experimental puzzles concerning $T_g$ shifts, the role of fragility, and mobility gradients of free standing polymer thin films that seem germane to our work. First, creep experiments \cite{56} suggest a universal correlation of the magnitude of $T_g$ shifts with polymer fragility is not valid. For example, PVAC has a high fragility of $m=135$ but very little reduction of $T_g$ is observed, while  PEMA has a much lower fragility of $m=87$ and a modest (but larger than PVAC) $T_g$ reduction \cite{56}. The PVAC behavior seems inconsistent with our present work, while the PEMA behavior is consistent. Indeed, older studies \cite{1,3} found that free standing films of PMMA and PS showed large quantitative differences in the magnitude of $T_g$ shifts (though not the functional form of the film thickness dependences), despite the fact that these two polymers having very similar bulk $T_g$ values, characteristic ratios, equation-of-state properties (e.g. cohesive energy), etc. Such large chemical variations remain largely a mystery from a microscopic theoretical perspective. They are not all captured by our statistical mechanical theory based on a minimalist model where polymer chains are mapped onto disconnected spherical Kuhn diameter segments.

A second puzzle is that although many experiments (and all simulations) find strong evidence for large mobility gradients, some calorimetric and creep experiments have been interpreted as not consistent with this deduction. For example, although the creep measurements discussed above do show large thin film $T_g$ shifts for both PS and PC \cite{56}, they do not seem to indicate a particularly large mobility gradient as evidenced by the time-thickness superposition behavior and by the observation that their glassy creep compliance is not higher than in the bulk. A possible resolution of the latter mystery has been suggested by Mirigian and Schweizer \cite{40} based on the idea that the creep measurements of the glassy compliance are not performed at the same fixed absolute temperature for films of different thicknesses.
\section{Discussion}
We have re-visited the ECNLE theory of glassy dynamics in free standing polymer thin films to improve the treatment of the collective elastic displacement field. Specifically, we go beyond the naive cut off of the isotropic bulk displacement field model employed previously \cite{38,39,40} to explicitly include some aspects of anisotropy and a modified boundary condition at the vapor interface. The consequences of this improvement are quantitative, not qualitative, but of significant magnitude and in the direction of increasing the speed up of dynamics due to a vapor interface and confinement. Semi-infinite thick films have also been studied for the first time, and the improved theory applied to address new questions for three different polymers of very different dynamic fragility.

The role of vitrification time scale was examined over a range of 9 decades of variability. Rather surprisingly, for the question of $T_g$ film-averaged shift normalized to its bulk value, the timescale criterion is found to have a relatively minor effect. This is good news for molecular dynamics (MD) computer simulation studies. As the vitrification criterion becomes shorter, absolute $T_g$ shifts do modestly decrease mainly because of the decreased importance of the longer range elastic barrier at the effectively higher temperatures. The mobile layer size was determined, and shown to grow strongly in the deeply supercooled regime. Moreover, the logarithm of the bulk isotropic liquid alpha time is predicted to be directly related to this film-defined length scale raised to a power modestly smaller than unity. We also showed that the theory predicts a new type of spatially inhomogeneous "dynamic decoupling" in films, corresponding to an effective factorization of the total barrier into its bulk temperature-dependent value multiplied by a function that depends only on location in the film. The corresponding decoupling exponent grows monotonically as the surface is approached, and bulk behavior (no decoupling) is not recovered until $\sim 15-20$ nm from the surface. Average thin film $T_g$ shifts were also studied as a function of film thickness and polymer chemistry. Larger shifts are predicted for psuedo-thermodynamic versus dynamic probes, for longer time scale vitrification criteria, and for more fragile polymers. Quantitative, no adjustable parameter comparisons with experiment and simulation for the thickness dependent shift are in reasonable agreement with the theory, including a nearly 100 $K$ suppression of $T_g$ in 4 nm PC films. Predictions were made for PIB, and to a lesser extent, PC films. 

Much remains to be done. First, it appears the theory does not make accurate predictions for the precise functional form of the barrier gradient in free standing thin and thick films, and hence by association the detailed spatial form of the alpha time and $T_g$ gradients. MD simulations performed in the dynamic crossover or lightly supercooled regime \cite{46,47,52} suggest a roughly double exponential variation of the alpha time as a function of distance from the vapor interface, in contrast to the theory. Hence, although it appears our theory with its minimalist treatment of multiple physical aspects can make good predictions for film-averaged properties, there seems to be missing physics. We suspect this involves another mechanism for enhancing mobility, likely mainly point (i) discussed in section III:  how surface-nucleated mobility "propagates" into the film. Our highly local approximation for this aspect may miss longer range surface-nucleated "facilitation-like" or "mobility-transfer" effects. This problem is under study. Beyond this, work continues on the open and difficult problem of how to treat the effects of solid surfaces or interfaces of variable mechanical stiffness on thin film (or bilayer, or droplet) glassy dynamics. 

\begin{acknowledgments}
This work was performed at the University of Illinois at Urbana-Champaign and was supported by DOE-BES under Grant No. DE-FG02-07ER46471 administered through the Frederick Seitz Materials Research Laboratory. We thank Professor David Simmons for many stimulating and informative discussions, for his suggestion to investigate decoupling effects in thin films with our theory, and for sharing his unpublished simulation results that discovered decoupling in thin and thick polymer films \cite{47,52}. 
\end{acknowledgments}

\end{document}